\documentstyle[aps,prl,twocolumn]{revtex}

\author{Dmitry V. Savin and Valentin V. Sokolov}
\address{Budker Institute of Nuclear Physics,
630090 Novosibirsk, Russia}
\title{ Quantum Versus Classical Decay Laws in Open Chaotic Systems}
\date{Phys. Rev. E 56, (1 Nov.1997) to appear as Rap. Comm.}

\begin{document}
\draft
\maketitle

\begin{abstract}
We study analytically the time evolution in decaying chaotic systems and
discuss in detail the hierarchy of characteristic time scales that appeared
in the quasiclassical region.  There exist two quantum time scales: the
Heisenberg time $t_H$ and the time $t_q=t_H/\sqrt{\kappa T}$ (with
$\kappa\gg 1$ and $T$  being the degree of resonance overlapping and the
transmission coefficient respectively) associated with the decay. If
$t_q<t_H$ the quantum deviation from the classical decay law starts at the
time $t_q$ and are due to the openness of the system. Under the opposite
condition quantum effects in intrinsic evolution begin to influence the
decay at the time $t_H$. In this case we establish the connection between
quantities which describe the time evolution in an open system and their
closed counterparts.
\end{abstract}

In a recent paper \cite{CMS97} G. Casati {\it et al.}, using numerical
simulations in the kicked rotor model with relaxation, have demonstrated
that a new time scale exists for a decaying quantum system in the deep
quasiclassical region.  After this time, which is much less than the
Heisenberg time $t_H=2\pi\rho$ (with $\rho$ being the mean level density 
and
$\hbar=1$), the decay law begins to deviate from the classical one.

The aim of the present paper is to show that such a time scale is, in fact,
a general feature of open quantum chaotic systems and is related to
peculiarities in fluctuations of the resonance widths. We describe these
fluctuations in the framework of the random matrix approach and employ the
formalism of the effective non-Hermitian Hamiltonian  which is commonly 
used
in the theory of resonance scattering \cite{VWZ,SZ89}.

Generally, decay properties of open quantum systems are related to
fluctuations in complex eigenvalues (resonance energies)
${\cal E}_n=E_n-(i/2)\Gamma_n$ of the effective Hamiltonian ${\cal H}$
via the two-point correlator of the Green's operator
${\cal G}({\cal E})=({\cal E}-{\cal H})^{-1}$. As typical examples one can
mention the $S$-matrix  \cite{VWZ,L77,DHM92,LSSS95a,PGR96} or time delay
\cite{SW92,Eck93,LSSS95b,FSS97,FS97} correlation functions. The simplest
quantity of such a kind is the leakage of the norm inside an open system
\begin{equation}\label{P}
 P(t) = (1/N)\left\langle\mbox{Tr}\left\{ \exp (i{\cal H}^{\dag}t)
              \exp (-i{\cal H}t) \right\}\right\rangle \, ,
\end{equation}
which is a somewhat simplified version of the decay functions considered
in Ref. \cite{DHM92}. Here, the angle brackets stand for the random matrix
ensemble average, and the equivalence of spectral and ensemble averages is
implied \cite{P79}.  It is easy to see that similar quantity (without the
ensemble averaging) has been numerically evaluated in \cite{CMS97}. In the
case of a closed system $P(t)\equiv1$ at any time. The time dependence in
eq.~(\ref{P}) appears due to antihermitian part of the operator ${\cal H}$.
The well-known relation between the time evolution operator
$\exp (-i{\cal H}t)$ and the Green's function ${\cal G}(E)$ enables one
to represent $P(t)$ in the form of the Fourier integral
\begin{equation}\label{Fourier}
 P(t) = \frac{1}{4\pi^2N} \!\!\!\int\limits^{\infty}_{-\infty}
 \!\!\! d\varepsilon e^{-i\varepsilon t}
 \!\!\!\int\limits^{\infty}_{-\infty}\!\!\! dE \!
 \left\langle\mbox{Tr}\, {\cal G}(E\!+\!\frac{\varepsilon}{2})
 {\cal G}^{\dag}(E\!-\!\frac{\varepsilon}{2})\right\rangle
\end{equation}
In the eigenbasis of the effective Hamiltonian the trace in
Eq.~(\ref{P}) can be represented in the form
\begin{eqnarray}
P(t) &=& \frac{1}{N}\biggl\langle\sum_{n,n'}U^2_{n'n}\,\exp
  \left\{-i({\cal E}_n-{\cal E}^*_{n'})t \right\}\biggr\rangle
                                                     \label{Peig} \\
 &=& \int\limits^{\infty}_{-\infty}
 \!\!\! d\varepsilon e^{-i\varepsilon t}
 \!\!\!\int\limits^{\infty}_{0}\!\! d\Gamma e^{- \Gamma t}
 R(\varepsilon,\Gamma)                               \label{Pint}
\end{eqnarray}
where $R(\varepsilon,\Gamma)$ denotes
$$
  R(\varepsilon,\Gamma) \! = \! \frac{1}{N}
 \biggl\langle \!\sum_{\mbox{\,}n,n'} U^2_{n'n}
 \delta\bigl[\varepsilon - (E_n \!-\! E_{n'})\bigr]
 \delta\bigl[\Gamma - \frac{ \Gamma_n \!+\! \Gamma_{n'} }{2}\bigr]
 \! \biggr\rangle
$$
and $U_{n'n}=\langle\psi_{n'}|\psi_n\rangle$ is the Bell-Steinberger
nonorthogonality matrix \cite{BS65} of the eigenvectors $|\psi_n\rangle$
of ${\cal H}$. This matrix differs from the unity only if resonances 
overlap.
It is worth noting that, contrary to the $\varepsilon$ dependence of the
function $R(\varepsilon,\Gamma)$ determined by the level spacings along the
real energy axis, its $\Gamma$ dependence is governed by widths themselves.
Therefore, the decay law can not be directly related to the distances
$\sqrt{(E_n \!-\! E_{n'})^2+(\Gamma_n \!-\! \Gamma_{n'})^2/4}$
between resonance levels in the complex energy plane. This is contrary to
what was conjectured in Ref.\cite{CMS97}.

Prior to exact calculating $P(t)$ we would like first to perform 
qualitative
analysis. As it will be justified below by the exact calculation, the main
features of $P(t)$ can be understood already from calculation of the
diagonal part
\begin{equation}\label{Pdiag}
P_d(t) = \frac{1}{N} \biggl\langle\sum_n e^{-\Gamma_n t} \biggr\rangle
 = \int\limits_0^{\infty}\!d\Gamma \, {\cal P}(\Gamma) \, e^{-\Gamma t} \ .
\end{equation}
Here we put approximately $U_{nn}=1$, neglecting its smooth dependence on
the index $n$. The function ${\cal P}(\Gamma)$ is the distribution of
the resonance widths. This function is explicitly known \cite{FS96,FS97} 
for
the case of the unitary ensemble which corresponds to the systems with
the broken time-reversal symmetry. Therefore, we use this ensemble to
demonstrate our general statements. In Ref.\cite{FS97} detailed analysis
of the width distribution has been performed, in particular, convenient
for our purposes integral representation
\begin{equation}\label{Pwidth}
 {\cal P}(\eta) = \frac{1}{\kappa\eta^2}\,\frac{1}{(M \!-\! 1)!}
 \int\limits_{\eta \kappa(1-T)/T}^{\eta \kappa/T}
 \!\!d\xi\ e^{-\xi+M\ln\xi}
\end{equation}
is given, with $\eta=\Gamma/\Gamma_W$ being the decay width
measured in the units of the Weisskopf width \cite{BW79,LW91}
\begin{equation}\label{GW}
\Gamma_W = MT/2\pi\rho \ ,
\end{equation}
where $T$ is the
transmission coefficient and the dimensionless parameter
$\kappa=2\pi\rho\Gamma_W=MT$ characterizes the degree of resonance
overlapping. The rate of small widths diminishes rapidly when the number
$M$ of (statistically equivalent) open decay channels grows.
For small overlapping, $\kappa\ll 1$, the density
${\cal P}(\eta)$ simplifies to the well-known $\chi^2_M$ distribution.
However, quasiclassics corresponds to $M\gg 1$ and strong overlapping
$\kappa\gg 1$ \cite{LW91}. In this case the width distribution decreases
exponentially at $\Gamma<\Gamma_W$ and follows the power law
$\sim(\Gamma_W/\Gamma)^2$ within some domain above $\Gamma_W$. In the
classical limit $M,\rho\rightarrow\infty$ but $\Gamma_W$ is kept fixed
and identified with the classical escape rate \cite{LW91}, an empty strip
appears below the value $\Gamma_W$ \cite{HILSS92,LSSS95a}.

Substituting Eq.~(\ref{Pwidth}) in Eq.~(\ref{Pdiag}) and changing the order
of integration, we come to the expression
\begin{equation}\label{Pmod}
 P_d(t) \!= \frac{1}{T}  \!\!\!\!\!\!  \int\limits_{0}^{{\ }\ T/(1-T)}
 \!\!\! \!\!\!  \frac{d\xi}{(1 \!+\! \xi)^2}\, \exp\left\{ \! -M\ln
 \bigl[1+\frac{1 \!+\! \xi}{M} \Gamma_W t\bigr] \! \right\}
\end{equation}
which is still exact in $M$ and $T$. In the classical limit defined above
the first term in the $1/M$ expansion of the logarithm in
Eq.~(\ref{Pmod}) gives $P_d(t)=P_{cl}(t)\,p(t)$, where
$P_{cl}(t)=\exp(-\Gamma_W t)$ is the classical decay probability which
follows from the semiclassical periodic orbit theory \cite{S91}
and $p(t)$ is a slowly varying factor, the proper calculation of which
lies beyond the diagonal approximation. Further terms of the $1/M$
expansion can be neglected for the times appreciably less than the
characteristic time
\begin{equation}\label{t_q}
t_q=\sqrt{M}\,t_W=\sqrt{\kappa/T}\,t_W =t_H/\sqrt{\kappa T}\ ,
\end{equation}
where $t_W\equiv 1/\Gamma_W$ is the characteristic lifetime of the system.
The quantum time scale $t_q$ is similar to that found in ref.\cite{CMS97}
(see also \cite{F97}). In the mesoscopic systems the typical life time is
given by the Thouless time \cite{AS86}. Therefore, the connection
$t_H=\kappa\, t_W$ shows that our overlapping parameter $\kappa$ plays the
role analogous to the dimensionless conductance in the mesoscopic physics.
We note in this respect that the ratio $t_W/t_q$ differs from that
conjectured in Ref.\cite{CMS97} by an additional factor $\sqrt{T}$,
which depends on the strength of coupling to channels. It is also
worth noting that the time $t_q$ appears in the relaxation phenomena
in disordered conductors as well \cite{relax,rem}.

The next-to-leading term of the expansion being positive, quantum 
corrections
slow down the decay law at $t>t_q$. After this time crossover occurs to
the asymptotic power law
\begin{equation}\label{P_as}
 P^{(as)}_d(t) = \kappa^{-1}\,\left( \Gamma_W t/M \right)^{-M} \, ,
\end{equation}
which is characteristic for open quantum systems \cite{LW91}. As it will
be demonstrated below, this expression correctly matches the exact result.
The fact that the diagonal approximation properly reproduces the
asymptotic behavior was first noted in \cite{ISS94}.
One can easily see from Eq.~(\ref{Pdiag}) that such a power behavior comes
from the influence of the widths which are smaller than $\Gamma_W$. Their
rate differs from zero as long as the parameter $1/M$ remains finite.

Qualitative arguments presented above can be put on a rigorous ground.
Powerful supersymmetry technique \cite{Efetov,VWZ} enables us to perform
exact calculations in Eq.~(\ref{Fourier}). Skipping the details of quite a
standard calculation, we concentrate on the analysis of the result.
For the case of unitary symmetry it reads
\begin{eqnarray}\label{super}
&&  P(t)= \!\int\limits_{-1}^1\!\!d\lambda_0 \!\!
 \int\limits_1^{\infty} \!\!d\lambda_1\,
 \mu(\lambda_i)\, f(\lambda_i)\,
 \delta(t/t_H - (\lambda_1\!-\!\lambda_0)/2)  \nonumber\\
&& \mbox{\quad\quad\quad} \times \left[ \frac{1\!+\!T(\lambda_0-1)/2}
 {1\!+\!T(\lambda_1-1)/2 }\right]^{M}
\end{eqnarray}
where $\mu(\lambda_i)= (\lambda_1\!-\!\lambda_0)^{-2}$ is
the measure of integration \cite{Efetov}
and $f(\lambda_i)=(\lambda_1^2-\lambda_0^2)/2$.
The openness of the system is contained in the last ``channel" factor in
Eq.~(\ref{super}). Actually, the structure of the expression (\ref{super})
is of universal nature for different quantities that describe the time
evolution of a chaotic quantum system. It consists of the integration with
the measure which is specific for the chosen ensemble, the channel factor,
and the preexponent $f(\lambda_i)$. The latter is the only factor that
depends on the concrete quantity considered. Since the time dependence
related to decay properties comes just from the channel factor, our 
analysis
is of quite a general meaning.

At $t<t_H$ the decay probability (\ref{super}) can be represented in
the form
\begin{eqnarray}\label{t<t_H}
 P(t) &=& \int\limits_{0}^{1}\!d\nu\, \left[1+(1-2\nu)t/t_H \right]  
\nonumber \\
 && \times
 \exp \left[ M \ln\left(1- \frac{\Gamma_W t/M}{1+(1-\nu)\Gamma_W t/M}
 \right) \right]\,,
\end{eqnarray}
whereas at $t>t_H$ it looks like
\begin{eqnarray}\label{t>t_H}
 P(t) &=& e^{ -M \ln (1+\Gamma_W t/M)}
 \!\int\limits_{0}^{1}\!d\nu\, \left[1+(1-2\nu)t_H/t \right]  \nonumber \\
 && \times
 \left[\frac{1-\nu T}{1 - \nu T/(1 +\Gamma_W t/M) } \right]^M\,.
\end{eqnarray}
In the classical limit (see above) simple calculation leads to the 
classical
decay law $P_{cl}(t)$.

The peculiarities of quantum deviation from the classical time evolution
depend significantly on the ratio $t_H/t_q=\sqrt{\kappa T}$.
If $\kappa T\gg 1$, so that $t_W\ll t_q\ll t_H$, the analysis of
Eq.~(\ref{t<t_H}) goes along the similar way as described below
Eq.~(\ref{Pmod}) and leads to the same conclusions. In the opposite case
$\kappa T\ll 1$ (but still $\kappa\gg 1$, which implies small values of the
transmission coefficient $T$) one arrives, by inspecting the integral 
factor
in Eq.~(\ref{t>t_H}), at the general relation
\begin{equation}\label{OC-fac}
P_{open}(t)=\left(1+\Gamma_W t/M\right)^{-M}\,P_{closed}(t)\,,
\end{equation}
which remains valid till the time
$t_{f}=t_W/T=t_q/\sqrt{\kappa T}=t_H/\kappa T$. Here $P_{closed}(t)$ is the
Fourier transform of the spectral correlation function which describes the
time evolution in the corresponding closed system.  Analogous calculation 
of
the function $P(t)$ for the case of the orthogonal ensemble (which
corresponds to chaotic systems with time-reversal symmetry) gives instead 
of
Eq.~(\ref{super}) an expression of similar structure but with obvious
changes which are characteristic to the symmetry class considered.  We
only mention that the channel factor contains the power $M/2$ rather than
$M$. Therefore, the same relation (\ref{OC-fac}) with $M$ substituted by
$M/2$ is valid also for time-reversal invariant systems.

The decay factor $\left(1+\Gamma_W t/M\right)^{-M}$ is equivalent to the
classical exponent till the time $t_q$. However, in the taken case the
Heisenberg time $t_H$ is smaller than $t_q$, which yields the influence of
quantum effects on the time evolution via the function $P_{closed}(t)$.  At
last, after the time $\kappa t_f=t_H/T$ the asymptotic regime
$P^{(as)}(t)\sim(\Gamma_W t/M)^{-M}$ appears with a proportionality
coefficient depending on the quantity considered. In the case of the
function (\ref{P}) its asymptotics coincides with that given by
Eq.~(\ref{P_as}).

In conclusion, even in the quasiclassical domain there exists a finite
probability of the widths less than the classical escape rate $\Gamma_W$.
This leads to the appearance of the new quantum time scale
$t_q=\sqrt{\kappa/T}\,t_W =t_H/\sqrt{\kappa T}$ associated with the decay.
The parameter of resonance overlapping $\kappa\gg 1$ plays the role
analogous to the dimensionless conductance in condensed matter physics.  
The
quantum effects begin to influence the time evolution starting from the 
time
$t_q$ if $t_q/t_H=1/\sqrt{\kappa T}\ll 1$ and from the Heisenberg time 
$t_H$
under the opposite condition. In the latter case the relation 
(\ref{OC-fac})
holds connecting the time evolution in an open system with its closed
counterpart.

We are grateful to D. Shepelyansky for interesting discussions, and thank
A.D.  Mirlin and V.G. Zelevinsky for useful remarks. Financial support from
INTAS Grant No. 94-2058 is acknowledged with thanks.

\end{document}